\documentclass{aa} 

\usepackage{graphicx}
\usepackage{txfonts}
\usepackage{subfig}

\usepackage{natbib}

\RequirePackage{xspace}

\def\ifundefined#1{\expandafter\ifx\csname#1\endcsname\relax}

\def\la{\mathrel{\hbox{\rlap{\hbox{\lower4pt\hbox{$\sim$}}}\hbox{$<$}}}}
\def\ga{\mathrel{\hbox{\rlap{\hbox{\lower4pt\hbox{$\sim$}}}\hbox{$>$}}}}

\newcommand{\be}{\begin{equation}}
\newcommand{\ee}{\end{equation}}
\newcommand{\bea}{\begin{eqnarray}}
\newcommand{\eea}{\end{eqnarray}}

\ifundefined{ensuremath}\def\ensuremath#1{\relax\ifmmode{#1}}
\else${#1}$\fi\else\relax\fi
\ifundefined{nuc}\def\nuc#1#2{\relax\ifmmode{}^{#1}{\protect\textrm{#2}}
\else${}^{#1}$#2\fi}\else\relax\fi

\newcommand{\nni}{\ensuremath{^{56}\mathrm{Ni}}\xspace}

\def\Teff{\ensuremath{T_{\mathrm{model}}}\xspace}

\def\tstd{\ensuremath{\tau_{\mathrm{std}}}\xspace}

\def\alog#1{\times 10^{#1}}
\newcommand{\phx}{\texttt{PHOENIX}\xspace}

\def\lstar{\ifmmode{\Lambda^*}\else\hbox{$\Lambda^*$}\fi} 
\def\Lstar{\ifmmode{\Lambda^*}\else\hbox{$\Lambda^*$}\fi} 
\def\Rop{\ifmmode{[R_{ij}]}\else\hbox{$[R_{ij}]$}\fi}

\def\Rji{\ifmmode{[R_{ji}]}\else\hbox{$[R_{ji}]$}\fi}
\def\Rstar{\ifmmode{[R_{ij}^*]}\else\hbox{$[R_{ij}^*]$}\fi}

\def\Rjistar{\ifmmode{[R_{ji}^*]}\else\hbox{$[R_{ji}^*]$}\fi}
\def\DRji{\ifmmode{[\Delta R_{ji}]}\else\hbox{$[\Delta R_{ji}]$}\fi}
\def\DRij{\ifmmode{[\Delta R_{ij}]}\else\hbox{$[\Delta R_{ij}]$}\fi}

\def\Jb{{\bar J}}
\def\alog#1{\times 10^{#1}}
\def\rin{\hbox{$R_{\rm in}$} }
\def\rout{\hbox{$R_{\rm out}$} }

\bibpunct{(}{)}{;}{a}{}{,} 

\begin{document}

   \title{A 3D radiative transfer framework: V. Homologous Flows}

   \author{E.~Baron
          \inst{1,2,3}
          \and
          Peter H. Hauschildt
          \inst{1}
          \and
          Bin Chen
          \inst{2}
          }

   \institute{Hamburger Sternwarte, Gojenbergsweg 112, 21029 Hamburg, Germany\\
              \email{yeti@hs.uni-hamburg.de}
         \and
             Homer L.~Dodge Department of Physics and Astronomy,
             University of Oklahoma, 440 W Brooks, Rm 100, Norman, OK
             73019-2061 USA\\ 
             \email{baron@ou.edu}
         \and
        Computational Research Division, Lawrence Berkeley
        National Laboratory, MS 50F-1650, 1 Cyclotron Rd, Berkeley, CA
        94720 USA
}

   \date{\today}

  \abstract
   {Observations and theoretical calculations have shown the
     importance of non-spherically symmetric structures in
     supernovae. Thus, the interpretation of observed supernova
     spectra requires the ability to solve the transfer equation in
     3-D moving atmospheres.}
   {We present an implementation of the solution of the radiative
     transfer equation in 3-D homologously expanding atmospheres in
     spherical coordinates. The
     implementation is exact to all orders in $v/c$.}
   {We use the methods that we have developed in previous papers in
     this series as well as a new affine method that makes use of the
     fact that photons travel on straight lines. The affine method
     greatly facilitates delineating the characteristics and can be
     used in the case of strong-gravitational  and arbitrary-velocity fields.}
   {We compare our results in 3-D for spherically symmetric test
     problems with high velocity 
   fields (up to 87\% of the speed of light and find excellent
   agreement, when the number of momentum space angles is high. 
    Our well-tested 1-D results are based on methods where
   the momentum directions vary along the characteristic (co-moving
   momentum directions). Thus, we are able to verify both the
   analytic framework and its numerical implementation. Additionally,
   we have been able to test the parallelization over
   characteristics. Using $512^2$ momentum angles we ran the code on
   16,384 Opteron processors and achieved excellent scaling.}
   {It is now possible to calculate synthetic spectra from realistic 3D hydro
     simulations. This should open an era of progress in hydro
     modeling, similar to that that occurred in the 1980s when 1-D
     models were confronted with synthetic spectra.}


   \maketitle

\section{Introduction}
\label{sec:01}

Supernovae of all types are known to deviate significantly from
spherical symmetry. The evidence comes from both flux spectra, but
particularly from the interpretation of spectropolarimetry
\citep[see][and references therein]{ww08araa}. In the case of
core-collapse supernovae, the asymmetry is thought to be due to the
underlying central engine which is probably asymmetric and this leads
to geometrically asymmetric ejecta, with the asymmetry growing as one
gets closer to the central engine (thus ``stripped'' supernovae such
as Type Ic are significantly more asymmetric than supernovae with
intact hydrogen envelopes such as Type IIP). Type Ia (thermonuclear)
supernovae are thought to be geometrically rather round but the
composition is thought to be asymmetrical. Since the light curve of
Type Ia supernovae is powered by the radioactive decay of \nni and its
products, asymmetries in the \nni distribution will lead to asymmetries
in the ionization fractions and opacities that will produce
polarization and alter the flux spectra. Thus, particularly in Type Ia
supernovae one can accurately calculate light curves and spectra
assuming homologous flow ($v \propto r$) but including the
geometrical or compositional asymmetry in three dimensions. 

In this series of papers \citep[][henceforth Papers
I-IV]{hb06,bh07,hb08,hb09a} we have built up the full characteristics 
method of solving the transfer equation in 3-D for static atmospheres
in a variety of geometries. Here we build on the results of Paper IV
for spherical geometry as well as those of \citet{bin07} for the
affine method.

\section{Transfer Equation}

\citet{bin07} showed that the transfer equation in flat spacetime
could be written in terms of an affine parameter and that the right
hand side could be evaluated in the co-moving frame provided that the
wavelength (or frequency) was evaluated in the co-moving
frame. However, the momentum directions could be held constant and
coincide with those of the observer's frame.

We define the rest frame  photon direction in spherical coordinates as
\be
{\bf n}=(1,\theta_n,\phi_n),\>   |{\bf n}|=1,
\ee or in Cartesian coordinates
\be
{\bf n}=(\sin\theta_n \cos\phi_n, \sin\theta_n\sin\phi_n, \cos\theta_n).
\ee and the starting position of the photon 
\be
{\bf r}_0=(r_0,\theta_0,\phi_0).
\ee
The $3$-D geodesic can be parametrized as
\be\label{r-s-eq}
{\bf r}(s)= {\bf r}_0+{\bf n}s,
\ee
where $s$ is the rest frame physical distance related to the affine
parameter $\xi$ by 
\be\label{s-xi}
s\equiv \frac{h}{\lambda_\infty}\xi,
\ee and is measured starting from ${\bf r}_0.$ This gives us
\be
\frac{dr}{ds}\equiv \dot{r}= \frac{ {\bf n} \cdot {\bf r}_0
  +s}{r}=\frac{{\bf n}\cdot{\bf r}}{r},
\ee
\be
\ddot{r}=\frac{1-\dot{r}^2}{r},
\ee and 
\be
r=|{\bf r}|=|{\bf r}_0+{\bf n}s|=\sqrt{r_0^2+2({\bf n\cdot r_0})s+s^2}, 
\ee
where
\bea
{\bf n\cdot r_0}&=& r_0[\sin\theta_0\sin\theta_n\cos(\phi_n-\phi_0)+\cos\theta_0\cos\theta_n],\cr
&=&x_0p_x+y_0p_y+z_0p_z.
\eea
Note also that
\be
\frac{d\gamma}{dr}\equiv \gamma'=\gamma^3\beta\beta',\>
\beta'=\frac{d\beta}{dr}. 
\ee
From Eqs.\,(\ref{s-xi}) above and (18) of \citet{bin07}, we find
\be\label{trans}
\frac{\partial I_\lambda}{\partial s}|_\lambda
+\frac{d\lambda}{ds}\frac{\partial I_\lambda}{\partial
  \lambda}=-(\chi_\lambda
\frac{\lambda_\infty}{\lambda}+\frac{5}{\lambda}\frac{d\lambda}{ds})I_\lambda+\eta_\lambda
\frac{\lambda_\infty}{\lambda} 
\ee
with
\be\label{lambda}
\frac{\lambda_\infty}{\lambda}=\gamma(r)\left[1-\dot{r}\beta(r)\right]\equiv
f(s). 
\ee
From Eq.\,(\ref{lambda}) we find
\bea\label{lambdadot}
\frac{1}{\lambda}\frac{d\lambda}{ds} &=& \frac{
  (\beta/r)\left(1-\dot{r}^2\right)-\gamma^2\beta'\dot{r}\left(\beta-\dot{r}\right)}{1-\dot{r}\beta(r)}\cr 
&\equiv& a(s).
\eea
Now we have
\be
\frac{\partial I_\lambda}{\partial s}|_\lambda + a(s)\lambda\frac{\partial I_\lambda}{\partial \lambda}=-[\chi_\lambda f(s)+5 a(s)]I_\lambda+\eta_\lambda f(s).
\ee
Finally, this can be put into the standard form used in \phx
\citep{hbmathgesel04,hbjcam99} 

\be
\frac{\partial I_\lambda}{\partial s} + a(s)\frac{\partial}{\partial\lambda}
(\lambda I_\lambda) + 4 a(s)I_{\lambda} = -\chi_\lambda f(s)
I_\lambda + \eta_\lambda f(s),
\label{eqn:phxform}
\ee
where $a(s)$ is given by Eq.\,(\ref{lambdadot}), and $f(s)$ is given by Eq.\,(\ref{lambda}).

In order to finite difference this equation we need to explicitly
difference the $\frac{\partial}{\partial\lambda}
(\lambda I_\lambda)$ term. As described in \citet{bin07} we can write
\be
\frac{\partial}{\partial\lambda}
(\lambda I_\lambda) = \frac{\lambda_l I_{\lambda_l} - \lambda_{l-1}
  I_{\lambda_{l-1}}}{\lambda_l - \lambda_{l-1}}
\ee

Then, Eq.~(\ref{eqn:phxform}) can be written as: 
\bea
\frac{dI_\lambda}{ds} &+&
\left[a(s)\frac{\lambda_l }
{\lambda_l - \lambda_{l-1}}  
+ 4 a(s)+\chi_\lambda f(s)\right]
I_\lambda\nonumber\\
&=& a(s)\frac{\lambda_{l-1} I_{\lambda_{l-1}}}
{\lambda_l - \lambda_{l-1}} +  \eta_\lambda f(s).
\label{eqn:phxform1}
\eea

Now we can define an effective optical depth,
\bea
d\tau &=&- \left( \chi_\lambda f(s) + 4a(s) +
  \frac{a(s)\lambda_l}{\lambda_l - \lambda_{l-1}} \right) ds\\
&\equiv& \hat\chi ds,
\eea
which defines $\hat\chi$. We can also define the traditional source
function $S_\lambda = \eta_\lambda/\chi_\lambda $, so that
Eq.~(\ref{eqn:phxform1}) becomes
\bea
\frac{dI_\lambda}{d\tau} &=& I_\lambda + \frac{\chi_\lambda}{\hat\chi_\lambda} \left(
S_\lambda f(s) + \frac{a(s)}{\chi_\lambda}\frac{\lambda_{l-1} I_{\lambda_{l-1}}}
{\lambda_l - \lambda_{l-1}}\right)\\
&\equiv&  I_\lambda + \hat S_\lambda,
\label{eqn:diffform}
\eea 
which defines $\hat S_\lambda$.

The more sophisticated discretization in $\lambda$ described in
\citet{hb04} or \citet{khb09} can also be implemented. For the case of
arbitrary velocity fields the method of \citet{khb09} will be
required. Nevertheless, this straightforward method yields excellent
results when compared with the more sophisticated treatment used in
the 1-D code (see below).

\section{Angular Integration}

To solve the scattering problem in the co-moving frame, we
need to calculate the mean intensity and $\lstar$ in the co-moving
frame. Recall that the specific intensity is calculated in a frame
where five of the six phase-space variables are actually observer's
frame quantities. In particular, the two momentum directions are fixed
observer's frame quantities. Thus, we need to perform the angular
integration in the co-moving frame.
We have 
\be
u= \gamma[1,{\vec\beta}],\>\> u_0 = [1,0,0,0],
\ee
and
\be
p = \frac{hc}{\lambda}[1,\mathbf{\hat n}]
.\ee  Then 
\[ d\Omega = \left(\frac{u_0\cdot p}{u\cdot p}\right)^2 d\Omega_0 \]
Now 
\be u\cdot p  = \frac{hc}{\lambda}\gamma [ 1 - \mathbf{\beta\cdot
  \hat n}],\>\>\> u_0\cdot p  = \frac{hc}{\lambda}. 
\ee Thus 
\bea
d\Omega &=& (\gamma[1 - \mathbf{\beta\cdot \hat n}])^{-2} d\Omega_0 \cr
&=&(\gamma[1 - \beta(r)\dot{r}])^{-2} d\Omega_0 \cr
 &=& f(s)^{-2}d\Omega_0. \label{domega}
\eea

\subsubsection{Computation of $\lstar$}

The computation of $\lstar$ proceeds following the same procedure as
in Paper I. As demonstrated by \citep{OAB} and \citep{ok87}, the
coefficients $\alpha$, 
$\beta$, and $\gamma$ can be used to construct diagonal and tri-diagonal
$\lstar$ operators for 1D radiation transport problems. In fact, up to the full
$\Lambda$ matrix can be constructed by a straightforward extension of the idea
\citep{hsb94,hb04}. These non-local $\lstar$ operators not only lead to
excellent convergence rates but they avoid the problem of false
convergence that is inherent in the $\Lambda$ iteration method and can also be
an issue for diagonal (purely local) $\lstar$ operators. Therefore, it is
highly desirable to implement a non-local $\lstar$ for the 3D case.  The
tri-diagonal operator in the 1D case is simply a nearest neighbor $\lstar$ that
considers the interaction of a point with its two direct neighbors. In the 3D
case, the nearest neighbor $\lstar$ considers the interaction of a voxel with
the (up to) $3^3-1=26$ surrounding voxels (this definition considers a
somewhat larger 
range of voxels than a strictly face-centered view of just 6 nearest neighbors).
This means that the non-local $\lstar$ requires the storage of 27
(26 surrounding voxels plus local, i.e., diagonal effects) times the total
number of voxels $\lstar$ elements.

The construction of the $\lstar$ operator proceeds in the same way as
discussed 
in \citet{phhs392} and Paper I. In the 3D case, the
`previous' and `next' voxels along each characteristic must be known so that
the contributions can be attributed to the correct voxel. Therefore, we use a
data structure that attaches to each voxel its effects on its neighbors.
The scheme can be extended trivially to include longer range interactions
for better convergence rates (in particular on larger voxel grids). However,
the memory requirements to simply store $\lstar$ ultimately scales like
$n^3$ where $n$ is the total number of voxels. The storage requirements can be
reduced by, e.g., using $\Lstar$'s of different widths for different voxels.
Storage requirements are not so much a problem if a domain decomposition
parallelization method is used and enough processors are available. 

We describe here the general procedure of calculating the $\lstar$ with {\em
arbitrary} bandwidth, up to the full $\Lambda$-operator, for the method in
spherical symmetry \citep{hsb94}. The construction of the $\lstar$ is
described in \citep{ok87}, so that we here summarize the relevant
formulae.  In the 
method of \citep{ok87}, the elements of the row of $\lstar$ are
computed by setting the 
incident intensities (boundary conditions) to zero and setting
$S(i_x,i_y,i_z)=1$ for one voxel $(i_x,i_y,i_z)$ and performing a formal
solution analytically.

We describe the  construction of $\lstar$ using the example of a single
characteristic. The contributions to the $\Lstar$ at a voxel $j$ are given by
\vbox{\bea
     \Lambda_{i,j} = 0  &\quad {\rm for\ }  i<j-1  \\
     \Lambda_{j-1,j} = f(s_{j-1})\gamma_{j-1}   &\quad {\rm for\ }  i=j-1 \\
     \Lambda_{j,j} = \Lambda_{j-1,j}\exp(-\Delta\tau_{j-1}) +
f(s_j)\beta^k_{j}  
                                        &\quad{\rm for\ }  i=j  \\
     \Lambda_{j+1,j} = \Lambda_{j,j}\exp(-\Delta\tau_{j}) + 
f(s_{j+1})\alpha_{j+1} 
                                        &\quad{\rm for\ }  i=j+1  \\
     \Lambda_{i,j} = \Lambda_{i-1,j}\exp(-\Delta\tau_{i-1}) 
                                        &\quad{\rm for\ }  j+1 < i \label{wide}
\eea}
These contributions are computed along a characteristic, here $i$ labels the
voxels {\em along} the characteristic under consideration. These contributions
are integrated over solid angle with the same method (either deterministic or
through the Monte-Carlo integration) that is used for the computation of the
$J$. For a nearest neighbor $\lstar$, the process of Eq.~\ref{wide} is stopped
with $i=j+1$, otherwise it is continued until the required bandwidth has been
reached (or the characteristic has reached an outermost voxel and terminates).
Comparing with the results of Paper I, the $\lstar$ operator is
altered \emph{simply} by the Doppler-shift factor $f(s)$ at the
appropriate point.

\section{From Co-moving Frame to Global Inertial Frame}
The specific intensity $I_\lambda$ is observer dependent, it is related to the observer invariant phase space distribution $F(x,p)$ by
\be
I_{\lambda}=-\frac{c^2}{h}(u\cdot p)^5F(x,p),
\ee therefore, the invariant quantity should be $I_\lambda\lambda^5.$ We immediately have
\be
I_{\lambda_\infty}=\left(\frac{\lambda}{\lambda_\infty}\right)^5 I_\lambda=\frac{I_\lambda}{f(s)^5}.
\ee We do not need to transform the direction vector, because when we
write down our transfer equation, the only co-moving quantity we used
is the co-moving wavelength, the other two momentum space variables
(e.g., $n_x$,$n_y$)are global inertial (for the case we are working
on, the $\bf n$ vector is the direction of photon in physical space,
not the direction seen by the co-moving observer).  For our flat
spacetime case, if we want the direction of photon seen by
$u^a=\gamma[1,{\bf \beta}],$ we simply need to do a Lorentz boost, for
example using equation 11.98 of \citet{jackson75}.

\section{Application examples}

As a first step we have built upon the MPI parallelized
Fortran 95 program described in Papers I-VI. The parallelization of
the formal solution is 
presently implemented over solid angle space as this is the simplest
parallelization option and also one of the most efficient (a domain
decomposition parallelization method will be discussed in a subsequent
paper). In addition, the Jordan solver of the Operator splitting
equations is parallelized with MPI.
The number of parallelization related
statements in the code is small.

Our basic continuum scattering test problem is similar to that discussed in
\citet{phhs392}, \citet{hb04} and Papers I-II.  This test problem covers a
large dynamic range of about 9 dex in the opacities and overall optical depth
steps along the characteristics and, in our experience, constitutes a
reasonably challenging setup for the radiative transfer code. The
application of the 3D code to `real' problems is in preparation and requires a
substantial amount of development work (in progress). 
Comparing this test case to real world problems in 1D we have found
that this test is 
close to the worst case scenario and that
convergence, etc is generally  better in real world problems. We use a
sphere with a 
grey continuum opacity parametrized by a power law in the continuum optical
depth $\tstd$. The basic model parameters are
\begin{enumerate}
\item Inner radius $\rin=10^{13}\,$cm, outer radius $\rout = 1.01\alog{15}\,$cm.
\item Minimum optical depth in the continuum
  $\tau_{\mathrm{std}}^{\mathrm{min}} =10^{-4}$ and maximum optical depth in
  the continuum $\tau_{\mathrm{std}}^{\mathrm{max}} = 10^{4}$.
\item Grey temperature structure with $\Teff=10^4$~K.
\item Outer boundary condition $I_{\rm bc}^{-} \equiv 0$ and diffusion
inner boundary condition for all wavelengths.
\item Continuum extinction $\chi_c = C/r^2$, with the constant $C$
fixed by the radius and optical depth grids.
\item Parametrized coherent \& isotropic continuum scattering by
defining
\begin{equation}
\chi_c = \epsilon_c \kappa_c + (1-\epsilon_c) \sigma_c
\end{equation}
with $0\le \epsilon_c \le 1$. 
$\kappa_c$ and $\sigma_c$ are the
continuum absorption and scattering coefficients.
\end{enumerate}
The test model is just 
an optically thick sphere put into the 3D grid.
This problem is used because the results can be directly compared
with the results obtained with  our 1D spherical radiation transport
code \citep{phhs392} to assess the accuracy of the method.
The sphere is centered at the center of the  Cartesian grid, which is 
in each axis 10\% larger than the radius of the sphere.
The
solid angle space was discretized in $(\theta,\phi)$ with 
$n_\theta=n_\phi$ if not stated otherwise. In the following 
we discuss the results of various tests. In all tests we use
the full characteristics method for the 3D RT solution.

\subsection{LTE tests}

\begin{figure}
\centering
   \includegraphics[width=0.65\hsize,angle=0]{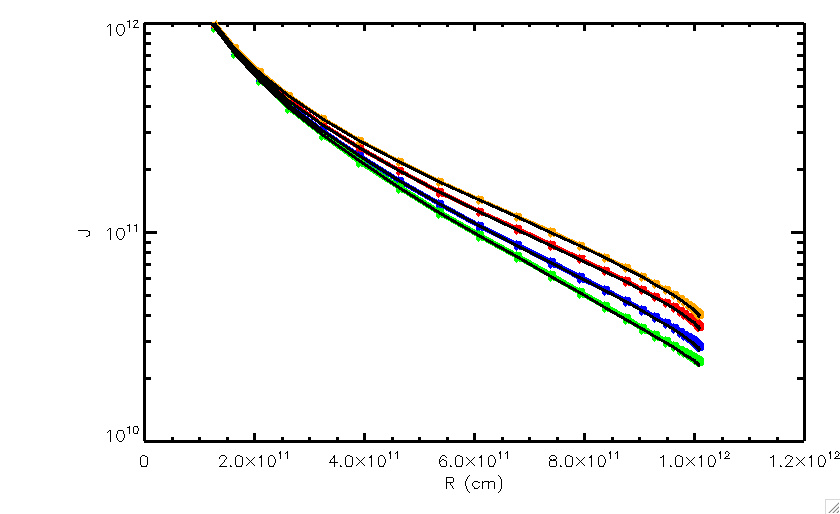}
      \caption{The results of 1-D calculations are compared with 3-D
   calculations for $\beta_{max} = (0.03,0.33,0.67,0.87)$. The
   momentum space directions were discretized using $n_\theta =
   n_{\phi} = 256$.}
         \label{fig:lte_high_v}
\end{figure}

In this test we have set $\epsilon=1$ to test the accuracy of the formal
solution by comparing to the results of the 1D code.  The 1D solver uses 64
radial points, distributed logarithmically in optical depth. For the
3D solver we tested `moderate' grids with $n_r=n_\phi=2*32+1$ and
$n_\theta = 2*16+1$ points along each
axis, for a total of $65^2*33 \approx 1.4\alog{5}$ voxels.
The momentum space discretization uses
in general, $n_\theta=n_\phi=256$ points. In
Fig.~\ref{fig:lte_high_v} we show the mean intensities as a function of
distance from the center for both the 1D ($+$ symbols) and the 3D
solver. For the 3D results $J$ is plotted at every voxel on the
surface and the spherical symmetry is reproduced very well at every
point on the surface. The
results show excellent agreement between the two solutions, thus the
3D RT formal solution is comparable in accuracy to the 1D formal
solution.  Fig.~\ref{fig:lte_high_v} shows the results for four
different maximum velocities, and clearly indicates that the 3D RT
solution is just as accurate as the 1D code. This is actually quite a
nice demonstration since the affine method used in the 3D code is
completely different from the full co-moving momentum space method
used in the 1-D code, that is, the specific intensity is solved for in
\emph{different} frames in the two codes. However, since $J$ depends
only on the co-moving frequency they can be directly compared in the
same frame.

As shown in Paper I, for the conditions used in these tests a larger
number of solid angle points significantly improves the accuracy of the mean
intensities. Our tests show that reasonable accuracy can be achieved
with as few as $16^2$ momentum space points, but in these test
calculations we have used more points in order to really compare the
3D results to the 1D results. \textbf{A full investigation of the
  number of angle points needed for realistic asymmetric calculations
  will be the subject of future work}.

The line of the simple 2-level model atom is parametrized by the ratio of the
profile averaged  line opacity $\chi_l$ to the continuum opacity $\chi_c$ and
the line thermalization parameter $\epsilon_l$. For the test cases presented
below, we have used $\epsilon_c=1$ and a constant temperature and thus a
constant thermal part of the source function for simplicity (and to save
computing time) and set $\chi_l/\chi_c = 10^2$ to simulate a strong line, with
$\epsilon_l = (1.0,0.1)$ (see below). With this setup, the optical depths as seen
in the line range from $10^{-2}$ to $10^6$. We use 92 wavelength points to model
the full line profile, including wavelengths outside the line for the
continuum.  We did not require the line to thermalize at the center of the test
configurations, this is a typical situation one encounters in a full 3D
configurations as the location (or even existence) of the
thermalization depths becomes more 
ambiguous than in the 1D case.

The sphere is put at the center of the  Cartesian grid, which is 
in each axis 10\% larger than the radius of the sphere.
For the test calculations we use voxel grids with the same
number of spatial points in each direction (see below). The
solid angle space was discretized in $(\theta,\phi)$ with 
$n_\theta=n_\phi$ Unless otherwise stated,
the tests were run on parallel computers using a variety of number of
CPUs, architectures, and interconnects.

\subsection{LTE line and continuum test}

We first have set $\epsilon_l=1$ to test the accuracy of the formal
solution by comparing to the results of the 1D code.  The 1D solver uses 64
radial points, distributed logarithmically in optical depth.  
Comparing the line mean intensities $\Jb$ as function
of distance from the center for both the 1D  and the 3D solver we
again found excellent agreement.

\subsection{Tests with line scattering}
\begin{figure}
\centering
   \includegraphics[width=0.65\hsize,angle=0]{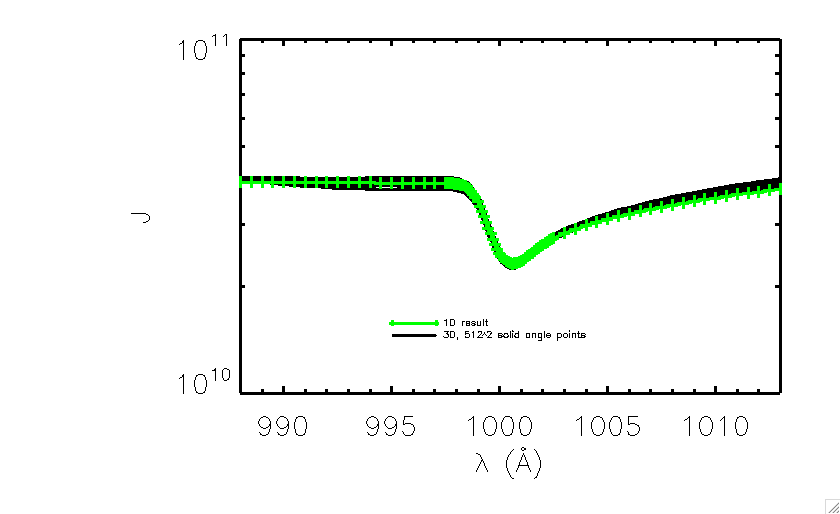}
      \caption{The results of 1-D calculations for a scattering line
   are compared with 3-D 
   calculations for $\beta_{max} = 0.03$. The
   momentum space directions were discretized using $n_\theta =
   n_\phi = 512$ and the calculation was run on the Franklin Cray
   XT4 using $2^{14} = 16384$ processors.}
         \label{fig:franklin_line}
\end{figure}

We have run a number of test calculations similar to the LTE case but with line
scattering included. In Fig.~\ref{fig:franklin_line} we show
the co-moving $J_\lambda$ as a function of $\lambda$ for
$\epsilon_l=0.1$ with $\beta_{max} = 0.03$ and $n_\theta = n_\phi =
512$.
The 3D calculations compare very
well to the 1D calculations. The small variation in some parts of the
surface of the sphere (each black line represents a pixel on the
surface of the sphere) is due to the different wavelength
discretization. The 3D case uses only the simple method described
above, whereas the 1D case uses the full Crank-Nicholson-like method
described in \citet{hb04}.

\begin{figure}
\centering
   \includegraphics[width=0.65\hsize,angle=180]{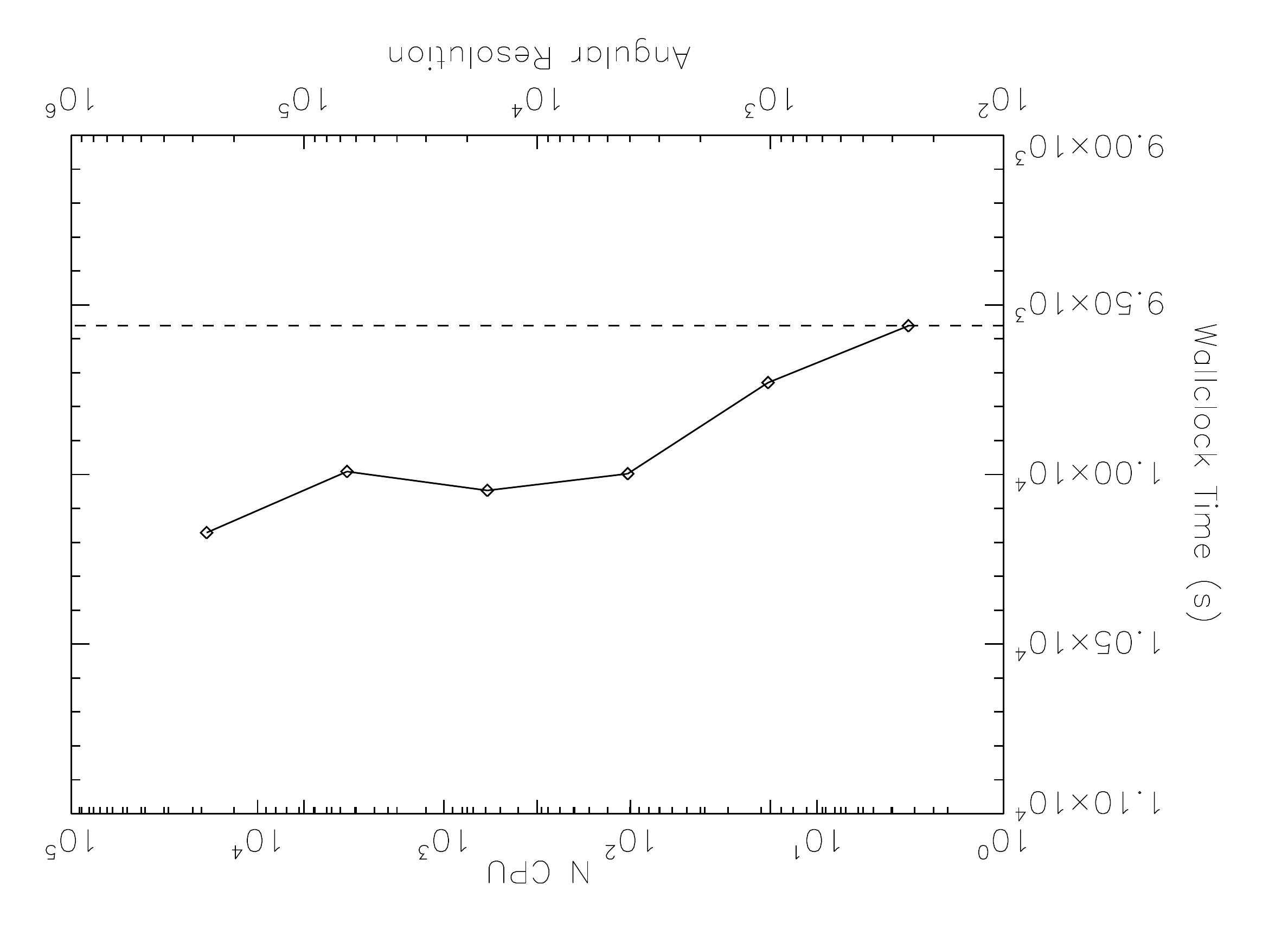}
      \caption{The wall-clock time to solve a scattering line $\epsilon
        = 0.1$ on 
   the Franklin Cray
   XT4 as a function of momentum frame angular resolution. The test
   was run so the amount of computational work per processor was
   constant. The roughly 14\% communication time increase from 16 to
   16,384 processors is acceptable.}
         \label{fig:speedup}
\end{figure}

Figure~\ref{fig:speedup} shows the wall-clock time as a function of
resolution (number of momentum-space angles $n_\theta$ and $n_\phi$ or
number of CPUs). The computational work is kept constant with each CPU
required to calculate 16 characteristics. The modest (14\%) increase
in wall-clock time from 16 CPUs to 16,384 CPUs is acceptable given the
huge increase in communication required and the fact that load
balancing is quite simple. \textbf{The wall-clock time for this test was about
625~s for each direction angle.  The memory usage is controlled by the
size of the spatial grid (this is also true in Papers I--IV). The only
additional storage is that the values of the Intensity for both the
previous
and current wavelength point must be stored. The total memory per
process was about 100 MB.} 

\section{Wavelength Parallelization}

We have implemented and tested a ``pipeline'' wavelength
parallelization method using wavelength clusters in the manner
described in \citet{bhpar298}. As in \citet{bhpar298}, the
parallelization over characteristics is done within a ``wavelength
cluster'' and each worker thus must send only its values of the
specific intensity to the corresponding process in the next wavelength
cluster. For the simple test problems considered
so far the opacity calculation is trivial and hence, there is no
speedup (but also no penalty) for this wavelength
parallelization. However, in real world problems the time to calculate
the opacities is roughly equal to the time required to solve the
transfer equation and this leads to linear speedups in the 1D case of
up to about a factor of eight.

\section{Conclusions}

We have implemented the affine method described in \citet{bin07} for
the case of homologous flows and
shown that it gives excellent results compared to the full co-moving
method that we use in our 1D code. We have also been able to
parallelize it both over characteristic and wavelength. The
characteristic scaling is excellent and immediately brings us into the
forefront of  massively parallel computation. The next step is to
include the 3D calculations in the full real world code and begin
applying it to numerous astrophysical problems.

\begin{acknowledgements} 
This work was supported in part  by SFB 676
from the DFG, NASA grant NAG5-12127,  NSF grant AST-0707704, and US DOE Grant
DE-FG02-07ER41517.    This research used resources of the National Energy
Research Scientific Computing Center (NERSC), which is supported by the Office
of Science of the U.S.  Department of Energy under Contract No.
DE-AC02-05CH11231; and the H\"ochstleistungs Rechenzentrum Nord (HLRN).  We
thank all these institutions for a generous allocation of computer time.
\end{acknowledgements} 

\bibliography{apj-jour,refs,baron,sn1bc,sn1a,sn87a,snii,stars,rte,cosmology,gals,agn,crossrefs}

\begin{appendix}
\section{Determining $s$ from Coordinates}

The characteristics are followed through the voxel grid from an
entry  boundary point to  an exit boundary point. It is convenient
to choose some particular voxel as a ``starting point'' $r_0$ and determine
the distance $s$ to the two boundary points. We will choose our sign
convention such 
that the distance to the entry point is negative and the distance to
the exit point is positive.
Given a starting point $r_0$ we can find the distance to a boundary
point $R$ (where $R=R_{\rm in}$, or $R=R_{\rm out}$) from
Eq.\,(\ref{r-s-eq}) and find 
\be\label{s-r-eq}
s^2+2({\bf n \cdot r_0})s+(r_0^2-R^2)=0
\ee
The characteristics can be divided into three classes. Tangential
characteristics (those that do not hit the inner boundary $R_{in}$ have 
\[
R=R_{\rm out},
\]  and satisfy the constraint that the impact parameter is greater
than $R_{\rm in}$, 
\be
 r_0^2-({\bf n\cdot r_0})^2>R^2_{\rm in}.
\ee For this case, Eq.\,(\ref{s-r-eq}) has two solutions
\be
s_-=-({\bf n \cdot r_0})-\sqrt{({\bf n \cdot r_0})^2-(r_0^2-R^2)},
\ee and
\be
s_+=-({\bf n \cdot r_0})+\sqrt{({\bf n \cdot r_0})^2-(r_0^2-R^2)},
\ee such that
\be
 s_-\leq 0\leq s_+.
\ee  

Core-intersecting characteristics include two cases, incoming and
outgoing rays. Incoming core-intersecting characteristics are
determined by
\be
{\bf n\cdot r_0}<0,
\ee where 
\[
R=R_{\rm out},
\] and there is only one solution 
\be
s_-=-({\bf n \cdot r_0})-\sqrt{({\bf n \cdot r_0})^2-(r^2_0-R^2)},
\ee the other solution
\be
s_+=-({\bf n \cdot r_0})+\sqrt{({\bf n \cdot r_0})^2-(r^2_0-R^2)},
\ee should be dropped because it passes through the core. Here
\be
s_-<0<s_+. 
\ee For outgoing core-intersecting characteristics
\be
{\bf n\cdot r_0}>0.
\ee this time the characteristic should start from the core
\[
R=R_{\rm in},
\] For this case 
\be
s_-=-({\bf n \cdot r_0})+\sqrt{({\bf n \cdot r_0})^2-(r^2_0-R^2)},
\ee is the desired solution and the other solution
\be
s_+=-({\bf n \cdot r_0})-\sqrt{({\bf n \cdot r_0})^2-(r^2_0-R^2)}
\ee passes through the core. In this case, 
\be
0>s->s_+.
\ee 

\label{lastpage}
\end{appendix}

\end{document}